\documentstyle[twoside,fleqn,espcrc2,epsfig]{article}

\title{A lattice field theoretical model for high-$T_c$ 
	superconductivity\thanks{Financially supported by DGICICYT AEN 96-16{70,74} and by Acci\'on Integrada Hispano-Francesa HF1996-0022.
        Presented by VMM (victor@lattice.fis.ucm.es).}}

\author{J.~L.~Alonso, \address{Departamento de F\'\i sica Te\'orica, 
	Universidad de Zaragoza, 50009 Zaragoza, Spain}
	Ph.~Boucaud, \address{LPTHE, Universit\'e de Paris XI, 91405 Orsay Cedex, France}
	V.~Mart\'{\i}n-Mayor\address{Departamento de F\'{\i}sica Te\'orica I,  
        Universidad Complutense de Madrid, 28040 Madrid, Spain}
	and A.~J.~van der Sijs\address{Swiss Center for Scientific Computing, 
	ETH-Z\"urich, ETH-Zentrum, CH-8092 Z\"urich, Switzerland}
        }

\begin{document}

\begin{abstract}
We present a 2+1-dimensional lattice model for the copper 
oxide superconductors and their parent compounds, in which both
the charge and spin degrees of freedom are treated dynamically.
The spin-charge coupling parameter is associated to the doping fraction 
in the cuprates. The model is able to account for the various phases of the
cuprates and their properties, not only at low and intermediate doping but
also for (highly) over-doped compounds. We acquire a qualitative understanding 
of high-$T_c$ superconductivity as a Bose-Einstein condensation of bound charge pairs.
\end{abstract}

\maketitle

\section{THE QUESTIONS}

The discovery of the new perovskite superconductors \cite{WM} opened the 
path to
critical temperatures ($T_c$) for superconductivity as high as $140$
K, surpassing by a factor of seven the highest known $T_c$ for metals
or alloys. In fact the physical mechanism for superconductivity seems
to be rather different. Traditional superconductors present an {\em
isotope effect}, which calls for a phonon mediated mechanism. They
also have a gap (of order of meV) in the excitation spectrum, as
shown by the exponential decrease of the specific heat at low temperature. 
All this can be accounted for by the BCS theory,
where electrons of opposite spin and momenta are attractively coupled
through a phonon exchange. The basic BCS statement is the existence
of a critical  temperature at which this attraction {\em
simultaneously} produces the {\em formation} of bound states of pairs of
electrons (with binding energy of the order of the gap) and the
{\em condensation} of those pairs into a {\it quantum liquid}.

The new superconductors show an intriguing phenomenology:

\begin{itemize}
{\item Superconductivity appears in doped, ceramic materials, like
La$_2$CuO$_4$ doped to 
La$_{2-x}$Ba$_x$CuO$_4$,
starting at $x\sim 0.05$.}

{\item At $x$=0 it is an insulating antiferromagnet.}

{\item At large $x$ it may be a normal metal.}

{\item The doping suppresses the antiferromagnetic (AFM) correlation
between neighboring CuO$_2$ planes. However, large ({\hbox{$\xi\sim 10a$}})
in-plain correlations remain just above $T_c$. Also, transport 
phenomena occur mainly in the CuO$_2$ planes, so everything looks 
like a d=2 problem, with localized spins on the Cu and mobile
holes on the O ions.}

{\item Again the superconductivity is charge-2 in nature (there is a 
{\em pairing state}).}

{\item Anomalous {\it normal} ({\em i.e.}
non superconducting) state: there are experimental indications of
a {\em pseudo-gap phase} (pair formation {\it before} quantum liquid 
condensation).}
\end{itemize}

All this poses some essential questions \cite{PINES}: What is the
physical origin of the anomalous {\it normal} state? How can it be
characterized? What is the mechanism for high $T_c$ superconductivity?
And what is the {\em pairing state}?

\section{OUR MODEL}

The natural (lattice) field-theoretical approach is 
effective field theory. For instance, the long wavelength, 
low temperature (antiferro)magnetic 
excitations of the the parent compound (undoped
La$_2$CuO$_4$) are well described by the O($3$) Non-Linear
$\sigma$-model (NL$\sigma$M) \cite{CHAKRA}. The basic
ingredients of our proposal are:

\begin {itemize}
{\item Treat localized spins (at Cu), and charge-carriers (at O)
independently and, both, dynamically {\em i.e.}\  not hole moving on a 
{\em fixed} spin background.}

{\item Relativistic lattice field theory, where M.C. and Mean Field
calculations are {\it feasible}.}

{\item Try to 
catch the essential features of  electrons strongly coupled
to AFM spin backgrounds ({\em i.e.} not a microscopic 
model).}
\end{itemize}

We define a 2+1-dimensional O($3$) lattice model, with two 
flavors of {\bf naive} fermions:
\begin{eqnarray}
S&=&\frac{1}{2}\sum_{n,\mu,f}\bar\Psi_n^f
\sigma^{\mu}(\Psi_{n+\mu}^f
-\Psi_{n-\mu}^f)\\\nonumber
&-&k \sum_{n,\mu} {\vec\phi}_n
{\vec\phi}_{n+\mu}
+y \sum_n
\bar\Psi_n^{f'}({\vec\phi}_n\cdot{\vec\tau})_{f'f} \Psi_n^f,
\label{ACCION}
\end{eqnarray}
where ${\vec\phi}=(\phi_1,\phi_2,\phi_3)$, ${\vec\phi}^2=1$ and
$\bar\Psi,\Psi$ are two-component Dirac spinors in $d=2+1$. The
flavors {\em mimic} the two values of the electron {\em spin}. The
matrices $\sigma^{\mu},\tau_i$ are Pauli matrices. The {\hbox
{spin-fermion}} coupling $y$ will be related to the doping fraction
$x$. In the large-$y$ region, the change of variable
($\Psi\to\frac{1}{y}{\vec\phi}\cdot{\vec\tau}\Psi$,$\bar\Psi\to\bar\Psi$)
shows that the fermionic kinetic term is suppressed by a factor $1/y$.
Therefore, at both $y=0$ and $y=\infty$ we recover the NL$\sigma$M.
The {\it insulating} undoped material should be represented by the
$y\to\infty$ limit, as suggested by the fermion hopping suppression.
Thus small $x$ corresponds to large $y$ (a possible correspondence is 
$x\sim C^2/(C^2+y^2)$, $C$ being a constant).

\begin{figure}[htb]
\begin{center}
\leavevmode
\centering\epsfig{file=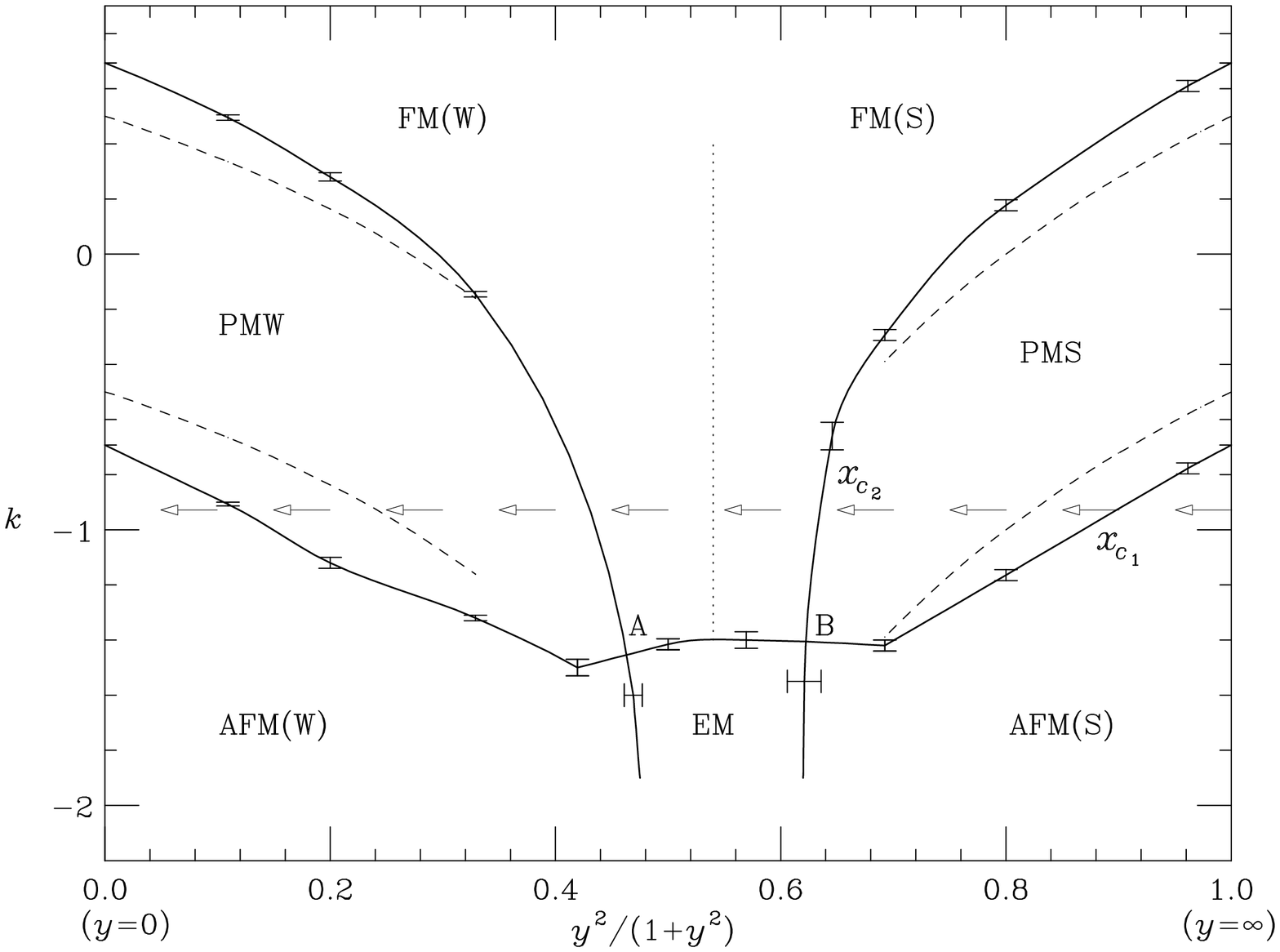,width=0.75\linewidth,angle=90}
\end{center}
\vglue -1cm
\protect\caption{Phase diagram of the action (\ref{ACCION}), with two
fermion families as required by HMC. Dashed lines indicates the MF calculation,
while solid are from a MC calculation on a $8^3$ lattice. 
\protect\label{PHASES}}
\end{figure}

There is a technical problem for the numerical simulation, as
${\vec\phi}$ are {\it constrained} variables not belonging to a {\it
Lie group}. We solve it by introducing {\it three} conjugate momenta
{\it per spin}. We then obtain the Hamiltonian
$$H=\frac{1}{2}\sum_n {\vec P}^2_n + S(\{ {\vec\phi}_n\}),$$
for which equations of motion preserving the constraint and the energy,
can be written:
$${\dot{\vec\phi}}_n={\vec P}_n\times{\vec\phi}_n\  ,\ 
{\dot{\vec P}}_n=-{\vec \phi}_n\times\frac{\delta S}{\delta{\vec\phi}_n}.$$
The standard HMC algorithm is now straightforward. The simulation
took 16 days of the 32 Pentium Pro processor parallel computer RTNN based in
Zaragoza. In fig.~\ref{PHASES}, we show the phase diagram of the model
at zero temperature. Notice that it is very similar to the familiar
diagram of  (chiral) Yukawa models \cite{YUK}. We name 
their ferrimagnetic phase {\it exotic magnetic} (EM), to avoid confusion in
a condensed matter context. So far, the
undoped material could be represented by either of
{\hbox{($\pm\kappa_0$,$y$=$\infty$)}} in
fig.~\ref{PHASES}. However, while doping the material, the
representative point moves in the plane ($\kappa$,$y$).
Doping suppresses the AFM order, so the curve $\kappa$=$\kappa$($y$)
should start at ($-\kappa_0$,$y=\infty$). In fig.~\ref{PHASES}, the
arrows sketch the possible {\em trajectory} of a material with
increasing doping.

For the fermionic correlation functions, our results are still at the
mean-field level. The small-$y$ region is in a perturbative
regime, with {\em light fermions} of mass
$m_f=y\langle{\vec\phi}\rangle$ (a Fermi liquid). In the
large-$y$ region, the fermionic kinetic term in the action is
{\hbox{$(1/y)\bar\Psi\gamma^{\mu}\partial_{\mu}\langle{\vec\phi}\cdot{\vec\tau}\rangle\Psi$}} at the MF level. Thus,
inside the PMS phase ($\langle{\vec\phi}\cdot{\vec\tau}\rangle = 0$), the fermions are essentially non-propagating. However,
the operator $\epsilon_{ff'}\Psi^f\Psi^{f'}$ excites a spin singlet of mass
$m^2_{pair}=4(y^2-3/2)$, which can get small in the PMS phase,
that is a {\em light charge-2 spin singlet bound state}, as in (chiral) Yukawa models \cite{STEPH}.  Inside  the
FM(S) phase, the kinetic term
is not so strongly suppressed ($\langle{\vec\phi}\rangle\neq 0$).
Thus, carriers are expected to propagate.
In the AFM(S) phase the original fermions (and doublers) do not
propagate. But fermions with
wavenumbers $\pm \pi/2$ are found to propagate (strongly resembling
the {\it Schrieffer pockets}). 

\begin{figure}[htb]
\begin{center}
\leavevmode
\centering\epsfig{file=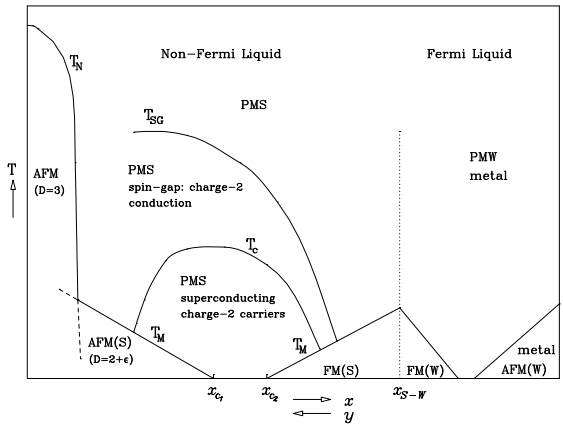,width=0.75\linewidth,angle=90}
\end{center}
\vglue -1cm
\protect\caption{Qualitative temperature ($T$) and doping ($x$) phase
diagram, from the model (\ref{ACCION}).
\protect\label{Tx}}
\end{figure}

Now, we can qualitatively discuss the behavior with the
temperature (see fig.~\ref{Tx}, and ref. \cite{NOS}). The $T$=0 axis can directly
be read off from the arrow line in fig.~\ref{PHASES}. As we enter
the PMS phase, we should expect the light, charge-2 bound states
to be Bose-Einstein condensed, yielding superconductivity. As 
the temperature rises, the BE condensation will disappear, and the system 
enters a region with heavy single fermions and uncondensed light 
charge-2 pairs. This we expect to correspond to the {\it pseudo-gap}
phase. At even higher temperatures, the pairs will break up (yielding
an insulating phase at the MF level). We also obtain Fermi liquid behavior
at large $x$ (small $y$). 

\section{OUR ANSWERS}

\begin{itemize}
{\item {\bf Physical origin of anomalous {\it normal} state?} A
dynamical, antiferromagnetically interacting spin background, {\it
strongly} coupled to fermions ({\it heavy} $\phi\Psi$ fermions)}.

{\item{\bf Characterization of anomalous {\it normal} state?} heavy single
charges ($\phi\Psi$) and {\it light  bosonic} charge-2 bound
states.} 

{\item {\bf Mechanism of high-$T_c$?} Bose-Einstein
condensation of previously formed {\it stable} pairs.}

{\item {\bf What is the pairing state?} A bound state of {\it heavy} fermions,
tied by spin-waves in a {\it disordered} phase: a {\bf PMS-pair}.}
\end{itemize}

Moreover, this pairing mechanism, as well as the non-coincidence of
pair formation and {\it quantum liquid} condensation, are likely to
occur in other spin-fermion models.


\begin{thebibliography}{99}

\bibitem{WM} J.G. Bednorz and K.A. M\"uller, 
{\sl Z. Phys.} {\bf B64} (1986) 184.

\bibitem{PINES} D. Pines, preprint cond-mat/9704102.

\bibitem{CHAKRA} S. Chakravarty, B.I. Halperin and D.R. Nelson,
{\sl Phys. Rev.} {\bf B39} (1989) 2344.

\bibitem{YUK} See, {\em e.g.} J. Shigemitsu,
{\sl Nucl. Phys.} (Proc. Suppl.) {\bf B20} (1991) 515.

\bibitem{STEPH} M.A. Stephanov, {\sl Phys. Lett.} {\bf B266} (1991) 447.

\bibitem{NOS} For further discussions and references, see
J.~L.~Alonso, Ph.~Boucaud, V.~Mart\'{\i}n-Mayor, A.~J.~van der Sijs,
preprint cond-mat/9706022 (and preprint SCSC-TR-97-16).
	

\end{thebibliography}
\end{document}